\newif\ifACM
\ACMfalse

\ifACM
\documentclass[format=sigconf, review=false]{acmart}
\else
\documentclass[10pt,conference]{IEEEtran}
\fi

\usepackage{graphicx}
\usepackage{xcolor}
\usepackage{listings}
\usepackage{comment}

\colorlet{light-gray}{gray!20}
\def\BibTeX{{\rm B\kern-.05em{\sc i\kern-.025em b}\kern-.08em
    T\kern-.1667em\lower.7ex\hbox{E}\kern-.125emX}}

\definecolor{dkgreen}{rgb}{0,0.6,0}
\definecolor{gray}{rgb}{0.5,0.5,0.5}
\definecolor{mauve}{rgb}{0.58,0,0.82}
\def\BibTeX{{\rm B\kern-.05em{\sc i\kern-.025em b}\kern-.08emT\kern-.1667em\lower.7ex\hbox{E}\kern-.125emX}}

\ifACM
\setcopyright{none}
\acmConference[ICPP '19]{ICPP '19: International Conference on Parallel Processing}{August 05, 2019}{Kyoto, Japan}

\renewcommand\footnotetextcopyrightpermission[1]{} 
\fi

\begin{document}

%
\title{A Framework for Model Search Across \\ Multiple Machine Learning Implementations}

\ifACM
\author{Yoshiki Takahashi}
\affiliation{%
  \institution{Tokyo Institute of Technology}
}
\email{yoshiki.takahashi.0326@gmail.com}

\author{Masato Asahara}
\authornote{Currently employed at dotData Inc. (March 2019).}
\affiliation{%
  \institution{NEC System Platform Research Laboratories}
}
\email{masahara@nec-labs.com}

\author{Kazuyuki Shudo}
\affiliation{%
  \institution{Tokyo Institute of Technology}
}
\email{shudo@is.titech.ac.jp}

\renewcommand{\shortauthors}{Yoshiki Takahashi, et al.}

\else

\author{
\IEEEauthorblockN{Yoshiki Takahashi}
\IEEEauthorblockA{
\textit{Tokyo Institute of Technology}\\
Tokyo, Japan
}
\and
\IEEEauthorblockN{Masato Asahara$^{\dagger}$}
\IEEEauthorblockA{
\textit{NEC System Platform Research Laboratories}\\
Kanagawa, Japan
}
\and
\IEEEauthorblockN{Kazuyuki Shudo}
\IEEEauthorblockA{
\textit{Tokyo Institute of Technology}\\
Tokyo, Japan
}
}
\fi

\ifACM
\begin{abstract}
Several recently devised machine learning (ML) algorithms have shown improved accuracy for various predictive problems.
Model searches,  which  explore  to  find  an  optimal  ML  algorithm  and hyperparameter values for the target problem, play a critical role in such  improvements.
During  a  model  search,  data  scientists typically  use  multiple  ML  implementations  to  construct  several predictive models; however, it takes significant time and effort to employ multiple ML implementations due to the need to learn how to use them, prepare input data in several  different formats, and compare their outputs.
Our proposed framework addresses these issues by providing simple and unified coding method.
It  has  been  designed  with  the  following  two attractive features: i) new machine learning implementations can be added easily via common interfaces between the framework and ML implementations and ii)  it can be scaled to handle large model  configuration  search  spaces  via  profile-based  scheduling.
The results of our evaluation  indicate that, with our framework, implementers need only write 55-144 lines of code to add a new ML implementation.
They  also  show  that  ours  was  the  fastest framework for the HIGGS dataset, and the second-fastest for the SECOM dataset.
\end{abstract}


\fi

\ifACM
\fi

%
\maketitle

\ifACM
\else

\fi

\renewcommand{\thefootnote}{\fnsymbol{footnote}}
\footnote[0]{$\dagger$ Current affiliation as of August 2019: dotData, Inc.}
\renewcommand{\thefootnote}{\arabic{footnote}}

\section{Introduction}
Machine learning (ML) techniques are frequently applied to
predictive analytics in a range of industries and academic fields, such as demand \cite{DemandPrediction}, price \cite{PricePrediction}, and click-through rate \cite{clickPrediction} prediction.
Modern ML implementations, such as XGBoost \cite{XGBoost}, TensorFlow \cite{TensorFlow}, and scikit-learn \cite{ScikitLearn}, enable us to achieve a higher accuracies than are possible with traditional ML algorithms, including linear regression and random forest.
The development of these high-performance ML implementations has accelerated the application of ML to various types of predictive analytics.

As the no-free-lunch theorem shows \cite{NoFreeLunchTheorem}, no single ML algorithm can achieve the lowest possible loss value for all loss functions, suggesting that we need to search for a different optimal ML algorithm for each prediction problem.
Data scientists usually try several different ML algorithms using multiple ML implementations, because each implementation supports several algorithms.
However, these types of model searches require a significant amount of time and effort on the scientists' part, implementations, prepare the data in suitable formats, and compare their outputs.

We propose a distributed framework for conducting model searches across multiple ML implementations.
Our framework has two particularly attractive features.
First, we can plug in a variety of ML implementations, even if they require data in different formats or programs written in different languages.
To achieve this, we employ common data formats and interfaces that conceal the differences between the framework and ML implementations.
Second, the framework can be scaled via parallelism.
We propose a profile-based scheduling approach that profiles the processing times of several training tasks and uses that information to schedule the tasks and assign them to workers.

Our proposed framework enables implementers to add a new ML implementation with only 55-144 lines of code, significantly less than would be required to implement such ML algorithms directly.
The results of our model search evaluation show that our framework was the fastest for the HIGGS dataset, and second-fastest for the SECOM dataset.

This paper is an extended version of our previous work \cite{SysMLWork,DataWorksSummit2018} limited to two pages, thus, could only give a brief description of the software's features and present some preliminary performance evaluation results.
Here, we use different datasets and frameworks for the comparison, yielding completely different results.

The remainder of the paper is organized as follows.
First, we review the background to model search and discuss our objectives in Section \ref{modelSearch}.
Then we describe our framework's design in Section \ref{frameworkDesign}, before discussing related work in Section \ref{relatedWork}.
Next, we present and evaluate our experimental results in Section \ref{evaluation}.
Finally, Section \ref{conclusion} concludes the paper.

\section{Model Search and Proposed Framework}\label{modelSearch}

Our aim is to obtain models that can accurately predict labels or target values for unseen future data. Such classification or regression problems can be handled by several ML techniques.
To find the most accurate predictive model for a given problem and dataset, we have to search for an optimal model, i.e., an optimal algorithm with the best possible hyperparameter values.

Usually, data scientists train their models using pre-existing ML implementations, such as XGBoost \cite{XGBoost}, TensorFlow \cite{TensorFlow}, or scikit-learn \cite{ScikitLearn}, instead of implementing the algorithm themselves.
Although scikit-learn \cite{ScikitLearn} and Spark MLlib \cite{Mllib} provide multiple ML algorithms, these do not perform as well as later implementations (e.g., XGBoost) for some ML algorithms. It is therefore important to use multiple ML implementations during the model search process.

That said, however, searching for a model across multiple ML implementations requires considerable effort on the part of data scientists.
First, they must learn how to use each implementation and read through the documentation for all its modules to code a model search program for each one.
Second, they must prepare the data in several formats, as different implementations require their own specific formats, such as row-oriented, column-oriented, or sparse-matrix formats.
Finally, they have to write a program to compare all the output models after training has been completed and select the best one.

Therefore, we propose a distributed framework that can search for a predictive model across multiple ML implementations.
This enables data scientists to execute and compare a wide range of
different algorithm implementations using a single data format and a unified coding method for describing the search space.
For example, they could use it to conduct model searches by running XGBoost and TensorFlow implementations in parallel, without having to consider their specific data formats and coding methods.

Our framework addresses two important challenges.
The first is designing the software so that new ML implementations can easily be added to the framework, while the other involves scheduling the training tasks, assigning them to execution units so that they all require approximately the same processing time.

\subsection{Simple and Unified Coding Method}

Our framework enables data scientists to use a single data format and a unified coding method to conduct model searches across multiple implementations, as shown in Figure \ref{figure:cord}.
Here, the first half of the code specifies the search space (possible ML implementations and hyperparameter values), while the second half performs the model search.
In this example, we explore 27 XGBoost configurations, 12 for TensorFlow, and 5 for scikit-learn.
Then, the second half of the program reads the training and validation data in a unified format and inputs them to the model search functions within the given search space.
This example shows how we can specify a model search in only 31 lines of code.

As well as simplifying the description of model search, we can also plug in a variety of ML implementations.
Although our framework is written in Scala, it can also incorporate implementations written in other languages.
For example, our evaluation (Section \ref{evaluation}) involved Python libraries for TensorFlow and scikit-learn.

\begin{figure}[t]
\centering\begin{minipage}{1.4\linewidth}
\begin{lstlisting}[linewidth=8cm]
/****** Search Space ********/
val xgbGrid = new GridBuilder()
 .addGrid(XGBoostParam.eta, Array(0.1,0.3,0.9))
 .addGrid(XGBoostParam.round, Array(30, 60, 90))
 .addGrid(XGBoostParam.maxBin, Array(32,64,128))
 .build
val tfGrid = new GridBuilder()
 .addGrid(TensorFlowParam.network,
 Array("128_128","64_64","128_64","64_64_64"))
 .addGrid(TensorFlowParam.learningRate,
 Array(0.003, 0.03, 0.3))
 .build
val sklearnLRGrid = new GridBuilder()
 .addGrid(ScikitLearnParam.algorithm,
 Array("logistic_regression"))
 .addGrid(ScikitLearnParam.c,
 Array(0.011, 0.033, 0.1, 0.3, 0.9))
 .build

/****** Model Search *******/
val trainDF = spark.read.load("../path/to/data")
val validateDF = spark.read.load("../path/to/data")
val searcher = new ModelSearcher()
 .addSpace(xgbGrid)
 .addSpace(tfGrid)
 .addSpace(sklearnLRGrid)
 .setFeaturesCol("features")
 .setLabelCol("label")
val multiModel = searcher.modelSearch(trainDF)
val scores = multiModel.validateAll(validateDF,
 "features", "label")
\end{lstlisting}
\end{minipage}
\caption{Sample Scala code for a model search using our framework.}
\label{figure:cord}
\end{figure}

\subsection{Adding New Implementations}
Our framework can easily be extended to incorporate new ML implementations.
Recent years have seen the rapid development of new ML algorithms, yielding improved accuracy for various prediction problems.
Thus, if we want to produce accurate predictions, we need to search for an optimal model within a space including these new algorithms.
Instead of implementing them ourselves, we should use third-party implementations for training.

However, there are several challenges involved in designing software that can flexibly incorporate multiple ML implementations.
One problem is the different formats used by each ML implementation to store the data.
These can include row-oriented, column-oriented, and sparse-matrix formats, using 32- or 64-bit values.
Thus, our framework must be able to handle multiple data structures.
Another problem is that interfacing with each implementation requires us to write a program in a specific language.
To scale our framework to handle multiple implementations, we overcome these problems by common interfaces described in Section \ref{interface}.

\subsection{Scaling up to Handle Large Search Spaces}\label{scaleOutProblem}
The space of model configurations covered during a model search can be very large, including various combinations of ML algorithms and hyperparameter values.
These have to be adjusted carefully, because the prediction accuracy can be very sensitive to the values used.
Since no one predictive model can be optimal for all predictive problems (by the no-free-lunch theorem), we need to search through several different algorithms.
Additionally, users may not evaluate predictive models based on accuracy alone, but also consider various other aspects, such as the execution time or the model's transparency, depending on the application.

In general, complex learning models, such as deep learning, are likely to achieve high accuracy, but if we focus more on transparency then simpler models, such as linear ones, may be preferable.
Thus, in order to find an optimal model based on several different evaluation metrics, it is often better to explore a variety of algorithms, from simple to more complex models.
Even with sophisticated parameter tuning techniques, such as Bayesian optimization \cite{BayesianOptimization}, it can take a very long time to explore such search spaces.

The time required to search through a large space can be reduced by parallel computing. Even though a single machine may take several days, or even several tens of days, to complete a search, we can accelerate this through parallelization, for example by dividing the search space into equal parts and assigning a separate machine to explore each part.
That said, it is difficult to reduce the processing time optimally simply by increasing the number of machines: since the training time can vary widely depending on the learning algorithm and hyperparameters used, the time taken by each machine to finish its exploration process can vary widely.
Thus, the slowest processing node dictates the total time required for the search and the performance improvement may not be ideal.
Section \ref{scheduling} describes our proposed profile-based scheduling solution.

\section{Framework Design}\label{frameworkDesign}
In designing our framework, our aim was to make it easy to incorporate multiple ML implementations and thereby find optimal predictive models within large search spaces.
This raised two issues.
The first is the different data formats and interfaces used by the implementations, making it difficult to include them in one unified framework.
The second issue is the difficulty of scheduling the model search in parallel across multiple machines to achieve good performance scaling.

In this section, we discuss the framework's overall architecture and the techniques used to address these issues.
It comprises four main modules, which we describe in Section 3.1.
Then, we show how the software is designed to make adding new ML implementations easy (Section \ref{interface}), and how we use profile-based scheduling to reduce the overall execution time (Section \ref{scheduling}).

\subsection{Architecture}\label{architecture}

\begin{figure}[t]
 \centerline {\includegraphics[bb=0 0 547.717 478.112,width=8cm]{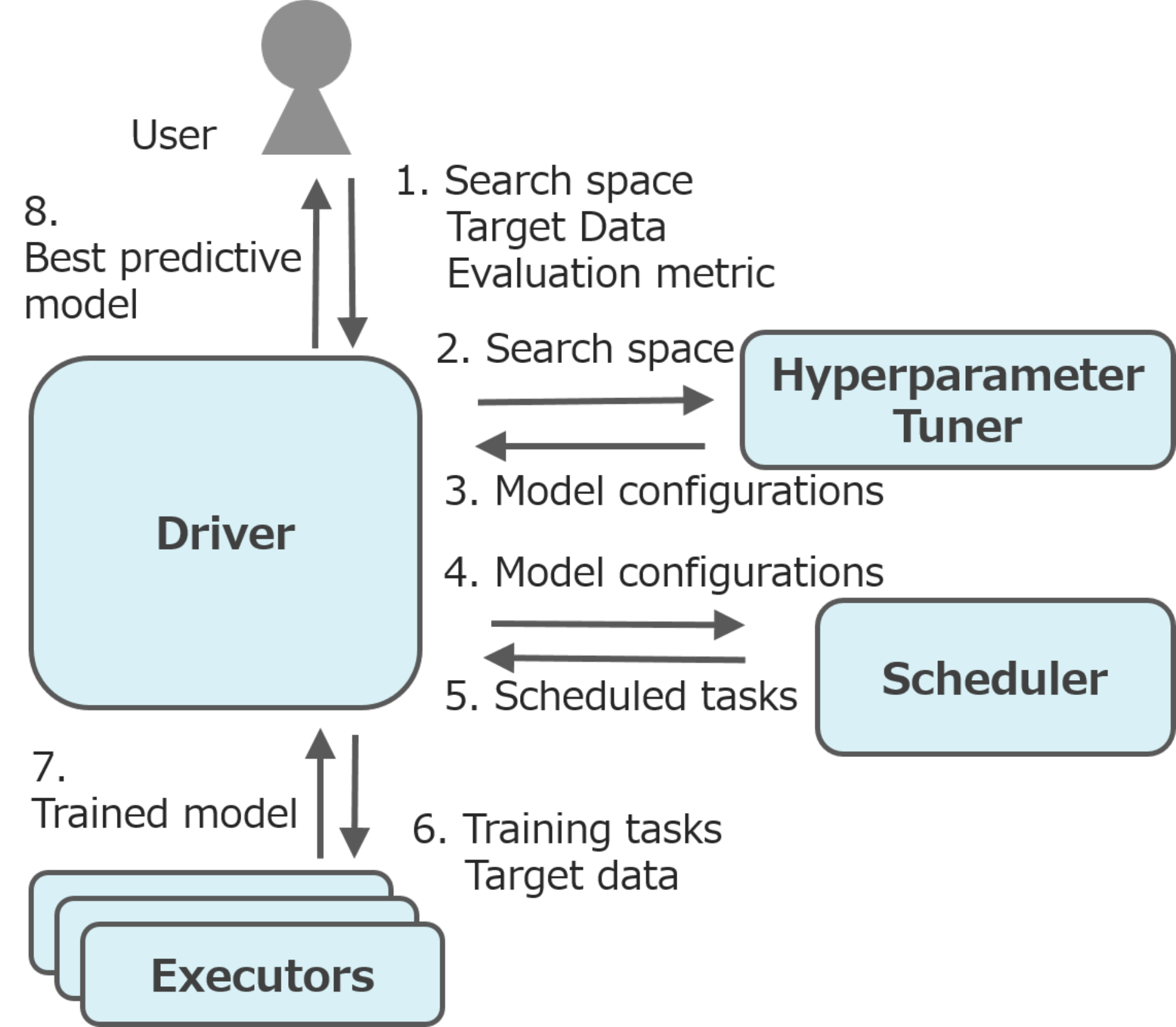}}
 \caption{
Architecture of our framework.
 }
\label{figure:architecture}
\end{figure}

Our framework comprises four main modules: {\it Driver}, {\it Hyperparameter Tuner}, {\it Scheduler} and several {\it Executors} (Fig. \ref{figure:architecture}).
Here, we discuss their functions during a model search.

First, the user provides a dataset, an evaluation metric (for validation), and a suggested search space to the {\it Driver}.
Then, the {\it Driver} passes the search space to the {\it Hyperparameter Tuner}, which returns a set of training configurations.
The {\it Hyperparameter Tuner} uses a static hyperparameter tuning algorithm, such as grid or random search \cite{RandomSearch}, to generate a set of model configurations.
Then, the {\it Driver} queries the {\it Scheduler} as to which {\it Executor} should execute which subset of configurations.
The {\it Scheduler} balances the load among the {\it Executors} by assigning them tasks based on profiling information (Section \ref{scheduling}).
The {\it Driver} then runs training tasks on all the {\it Executors} using the Apache Spark \cite{Spark} map function and obtains the trained predictive models.
These models are then evaluated by the {\it Executors}, similarly to the training process, and the best model (according to the given evaluation metric) is returned to the user.

Incorporating a new ML implementation into the framework involves making it runnable by the {\it Executors}.
In Section \ref{interface}, we discuss how the framework is designed to be extended, enabling ML implementations using different languages or data structures to work correctly on the {\it Executors}.
In Section \ref{scheduling}, we introduce the method adopted by the {\it Scheduler} to scale up the search to large search spaces.

\subsection{Common Interface to Hide Implementation Differences}\label{interface}
Every time a new learning algorithm is devised, the model search space must be extended.
Compared with implementing a new ML algorithm from scratch, simply writing glue code to reuse an existing algorithm implementation can drastically reduce the amount of additional code required.
However, plugging in a new implementation in this way requires the {\it Driver} to be modified to handle the differences between the implementations.

To simplify this, we have designed common interfaces, which the Driver uses to call training and prediction functions, which conceal the interface differences.
Since the {\it Driver} always uses the same interfaces to call these functions, implementers do not need to make any changes to it in order to plug in new implementations.
Instead, they simply define new training and prediction functions that inherit from these interfaces and invoke new third-party functions within them.
Additionally, there is no need to consider distributed processing because the training processes are executed locally on the {\it Executors}.

The gap of data formats is resolved on the {\it Executors}.
The common interface discussed above takes uniformed format data, which is represented as a row-oriented dense matrix.
This uniformed format data is going to be converted into a specific format for new implementation just before training.

The data format differences are resolved by the {\it Executors}.
The common interface discussed above takes data in a uniform format, represented as a row-oriented dense matrix.
This is then converted into the format needed by each implementation immediately prior to training.

Ideally, the common format should be determined based on the target data.
Tables containing mostly zero or empty values can be more heavily compressed by storing them in a sparse-matrix format instead of a dense format, while those containing few zero values will be poorly compressed by a sparse-matrix format.
Our framework does not support the variety of these formats yet but a dense matrix representation as a common format.

\subsection{Profile-based Scheduling}\label{scheduling}
It is difficult to optimally reduce the processing time by increasing the number of {\it Executors}. For example, dividing the search space into equal parts and assigning them to different {\it Executors} will not generally improve performance as much as expected, as different machines will have different processing times (Section \ref{scaleOutProblem}).

Here, we propose a scheduling method that uses profiling results to balance the task loads.
The overall processing time can be expressed by the maximum processing time over all nodes.
Assuming that all the task processing times are known, allocating the training tasks to nodes is an instance of an optimization problem known as job-shop-scheduling \cite{JobShopScheduling}.
Since this problem is NP-hard \cite{JobShopNPHard}, we solve it using an approximate (greedy) method.

To solve job-shop-scheduling, the framework must know all the task processing times.
First, it samples a few percent of the data from the whole dataset.
This is used for training, to profile the processing times of several training tasks in the search space.
Based on assuming that the training time is proportional to the data size, the {\it Scheduler} uses the profiling time divided by the sampling rate as the task processing time.

\begin{figure}[t]
 \centerline {\includegraphics[width=9cm]{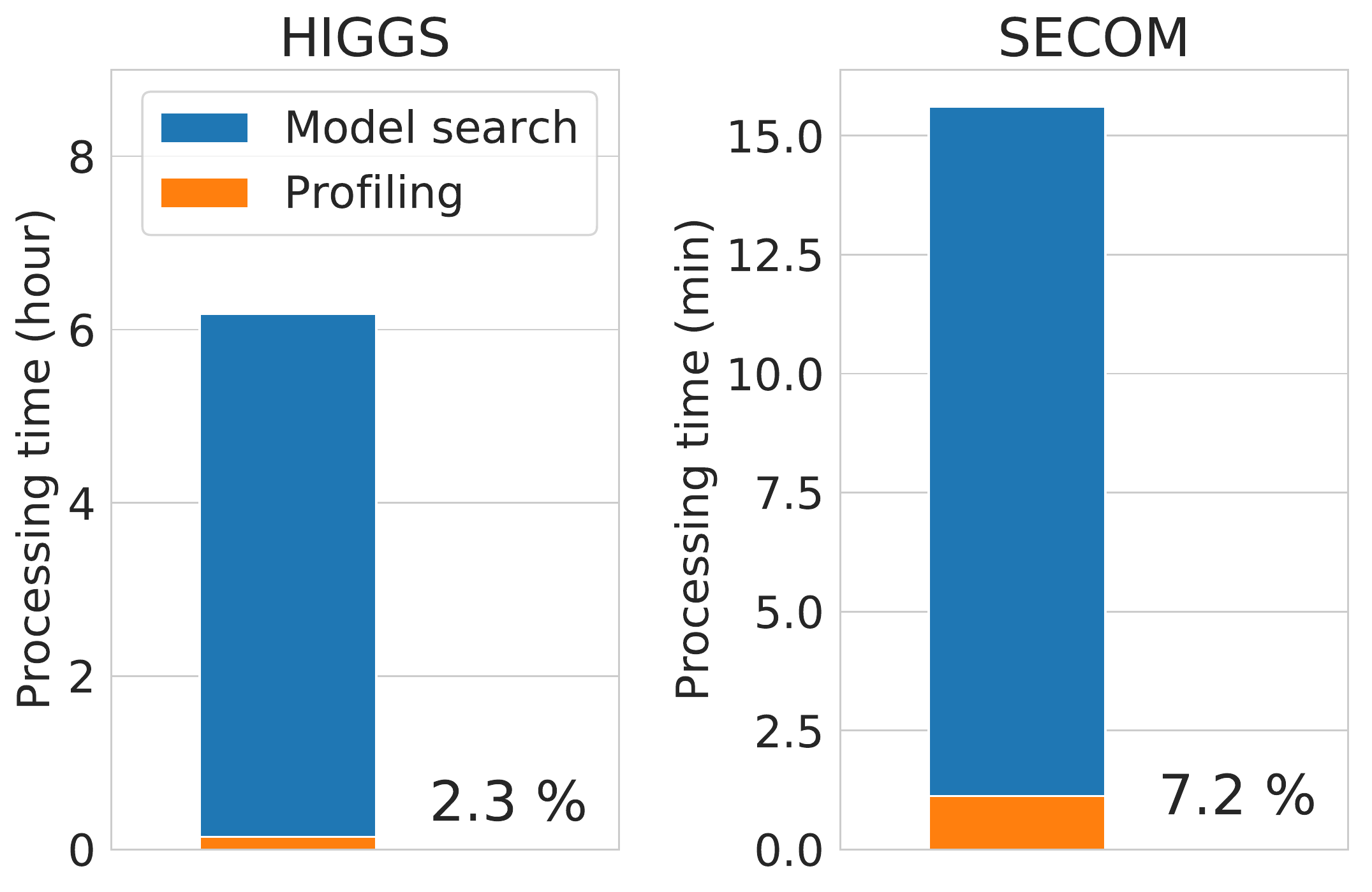}}
 \caption{
Ratio of profiling time to total processing time.
 }
\label{figure:profileRate}
\end{figure}

This scheduling method is useful when the time required for profiling is sufficiently short compared with the overall execution time.
Figure \ref{figure:profileRate} shows the ratio of the profiling time to the total processing time for the HIGGS and SECOM datasets (Section \ref{setUp}) based on exploring four ML algorithms and 1,211 model variants.
We sampled 1\% and 3\% of the data, respectively, to profile the datasets and solved the job shop scheduling problem using a greedy method.
The other settings were as as described in Section \ref{setUp}.
Here, profiling required less than 8\% of the overall execution time, which we believe is sufficiently fast.

Although we only discuss static scheduling here, it is also necessary to predict the processing time for dynamic scheduling.
In dynamic scheduling, once a worker has finished one job, it acquires the next one, and all nodes continue to operate until no unexecuted jobs remain.
Even with such dynamic scheduling, if the last job assigned to a worker is a long one, the other workers may have to wait for that worker to complete it, meaning we need to account for this in predicting the processing time.

\section{Related work}\label{relatedWork}
In this section, we discuss two topics related to model search: distributed model search frameworks, and hyperparameter tuning techniques.

\subsection{Distributed Model Search Frameworks}
There are several distributed frameworks that are designed to accelerate the time-consuming predictive model construction.
The Spark MLlib \cite{Mllib} framework aims to scale up model search performance to huge amounts of data.
It provides several ML algorithms and a function to automatically search for optimal predictive models based on hyperparameter values for each learning algorithm.
By default, model searches evaluate the model configurations serially, but it can be configured to operate in parallel.
However, since it trains a single model in parallel, the communication overhead caused by its shuffle process affects performance significantly more than with our framework.
Furthermore, it is difficult to extend existing ML implementations to run in parallel for use with MLlib.

The spark-sklearn \cite{SparkSklearn} framework runs scikit-learn's model search method on multi-node clusters with Apache Spark.
It can explore all ML algorithms supported by scikit-learn, and apply grid-search hyperparameter optimization.
Its scheduler divides training tasks into groups and assigns one group to each worker.
Our framework differs from spark-sklearn in that it can incorporate multiple ML implementations and use profile-based scheduling.

The MLbase \cite{MLbase} framework is designed to construct predictive models using distributed systems based on simple descriptions. It enables users to execute complicated steps, such as preprocessing, model verification, and model selection, without deep system knowledge.
TuPAQ \cite{AutomatingModelSearch} , building on MLbase, can obtain highly accurate predictive models in a short time using a scheduling method called bandit resource allocation that assigns computing resources to regions of the search space where more accurate models are likely to lie.
By contrast, our framework's scheduling approach aims to minimize the overall execution time required to complete all the training tasks.

The HyperDrive \cite{HyperDrive} framework's scheduling method has evolved over time into bandit resource allocation.
However, bandit resource allocation cannot cope with cases where the accuracies of two models reverse their order after several learning iterations.
To deal with this problem, HyperDrive's scheduling approach is based on the potential future prediction accuracy.

\subsection{Hyperparameter Tuning Algorithms}
Several proposed algorithms aim to construct more accurate predictive models by selecting which model configurations to explore within the search space.
Grid search and random search \cite{RandomSearch} are two classic hyperparameter tuning algorithms.
Grid search tries all the grid points between hyperparameter values in a pre-determined set.
Since the user can manually determine the hyperparameter values that are actually considered, this approach is suitable for analyzing the impact of particular hyperparameters on accuracy.
The random search method evaluates several sets of randomly selected values from the search ranges specified for each hyperparameter.

In contrast to these static methods, there are also several techniques for dynamically determining future hyperparameter values to consider based on previously evaluated configurations \cite{SMBO,BayesianOptimization,ExperiencedOpt,ReinforcementOpt}.
Bayesian optimization \cite{SMBO,BayesianOptimization} is frequently used for this, and has been adopted by several frameworks, such as Auto-WEKA \cite{AutoWEKA} and Hyperopt \cite{HyperOpt}.

Our framework can utilize these algorithms by incorporating them into its {\it Hyperparameter Tuner}.
This generates model configurations to evaluate based on a user-specified tuning algorithm.
When using dynamic tuning algorithms, it iteratively receives the evaluation results and uses them to generate new model configurations.

\section{evaluation}\label{evaluation}
Now that we have discussed the two issues our framework addresses, and their solutions, we investigate its performance experimentally.
First, to demonstrate that the framework can easily incorporate new ML implementations, we measure the changes in the number of lines of code after adding several third-party implementations.
Second, we perform model searches using the framework to compare its performance with those of existing frameworks.

\subsection{Experimental Setup}\label{setUp}
\paragraph{Workloads}
We used two binary classification datasets with numeric-valued attributes, taken from a UCI dataset repository \cite{UCI}.
The HIGGS dataset\footnote{https://archive.ics.uci.edu/ml/datasets/HIGGS} is derived from Monte Carlo simulations of physics events.
From this, we sampled 100,000 instances each for the training, validation, and testing sets and standardized them for use in model search.
The SECOM dataset\footnote{https://archive.ics.uci.edu/ml/datasets/secom} consists of signal data collected from sensors and process measurement points used in semiconductor manufacturing.
This included 1,567 instances, which we divided in a 6:2:2 ratio into training, validation, and testing sets, then standardized.

Using our framework, we executed model searches with 1, 2, 4, 8, 16, and 32 parallel tasks, measuring their execution times.
All the experiments were conducted on x86-64 servers, each with a single 4-core Intel Xeon E3-1280 CPU running at 3.70 GHz and 32 GB of memory.
We used one such machine for 1, 2, and 4 parallel tasks, and clusters of 2, 4, and 8 machines for 8, 16, and 32 tasks, respectively.

Here, we compared the results for two scheduling methods: random scheduling, which randomly assigns equal numbers of training tasks to all nodes, and the proposed profile-based scheduling method.
For profiling, we sampled 1\% and 3\% of the data for the HIGGS and SECOM datasets, respectively, and used the results to train models for all configurations in the search space.
The scheduler then solved the optimization problem using a greedy algorithm, estimating the processing time by dividing the training times by the sampling rate.

\paragraph{Model search space}
We considered four algorithms, exploring a total of 1,211 training configurations (algorithm-hyperparameter pairs) using grid search.
Gradient boosting tree is an ensemble learning algorithm that combines many decision trees.
For this algorithm, we explored 864 configurations by changing 6 hyperparameters, and executed it by running XGBoost \cite{XGBoost} inside the framework.
Multilayer perceptrons (MLPs) are artificial neural networks where each layer weights its input values and then applies nonlinear functions to produce output that it sends to the following layer.
Here, we used TensorFlow \cite{TensorFlow} to train the models using 324 configurations, changing factors such as the numbers of layers and neurons and the learning rate.
Finally, we also considered random forests and logistic regression, two traditional ML algorithms, which we trained using scikit-learn \cite{ScikitLearn}, using 18 and 5 configurations respectively.

\subsection{Comparison with Existing Frameworks}
We compared our framework with two existing frameworks: Spark MLlib and spark-sklearn.
Using these frameworks, we ran model searches with the same ML algorithms and hyperparameter values described in Section \ref{setUp}.
Spark MLlib \cite{Mllib} is a machine learning library, aimed at distributed large-scale data, which provides hyperparameter tuning and model selection functions, as well as several ML algorithms.
The spark-sklearn \cite{SparkSklearn} framework runs scikit-learn's model search on multi-node clusters with Apache Spark.
We compared these with our framework in terms of the execution time and the accuracy for model search.

Spark MLlib also has a parameter that enables several models to be evaluated in parallel during model search, although it evaluates the models serially by default.
In this evaluation, we investigated its performance for parameter values of 4-12 and selected 8 for that parameter.

\subsection{Results}

\begin{figure}[t]
 \includegraphics[width=8cm]{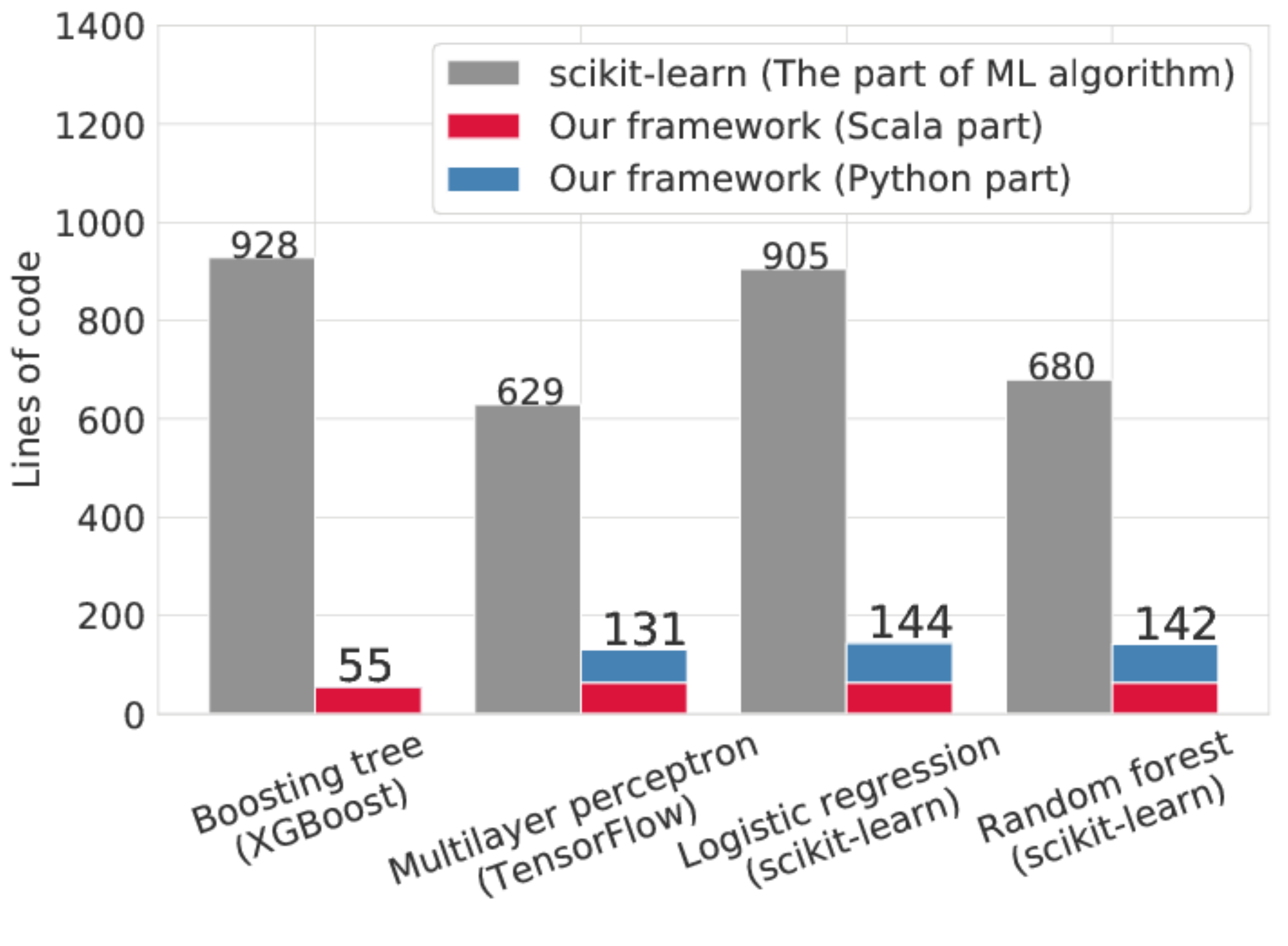}
 \caption{Numbers of code lines required to incorporate different ML algorithms into our framework.}
\label{figure:lines}
\end{figure}

Figure \ref{figure:lines} shows the number of lines of code required to add each new algorithm to our framework, with the red and blue bars showing the numbers of lines of Scala and Python code required, respectively.
This shows that we can include a new ML implementation with only 55-144 lines of code.
Although ML implementations that support the Java API (e.g., XGBoost) can easily be added to our framework because it is written in Scala, Python libraries, such as TensorFlow and scikit-learn, are more costly to incorporate.
Even so, compared with implementing each ML algorithm directly (gray bars), only needing to write glue code for our framework reduced the implementation cost drastically.

\begin{figure}[t]
 \centerline {\includegraphics[width=8.3cm]{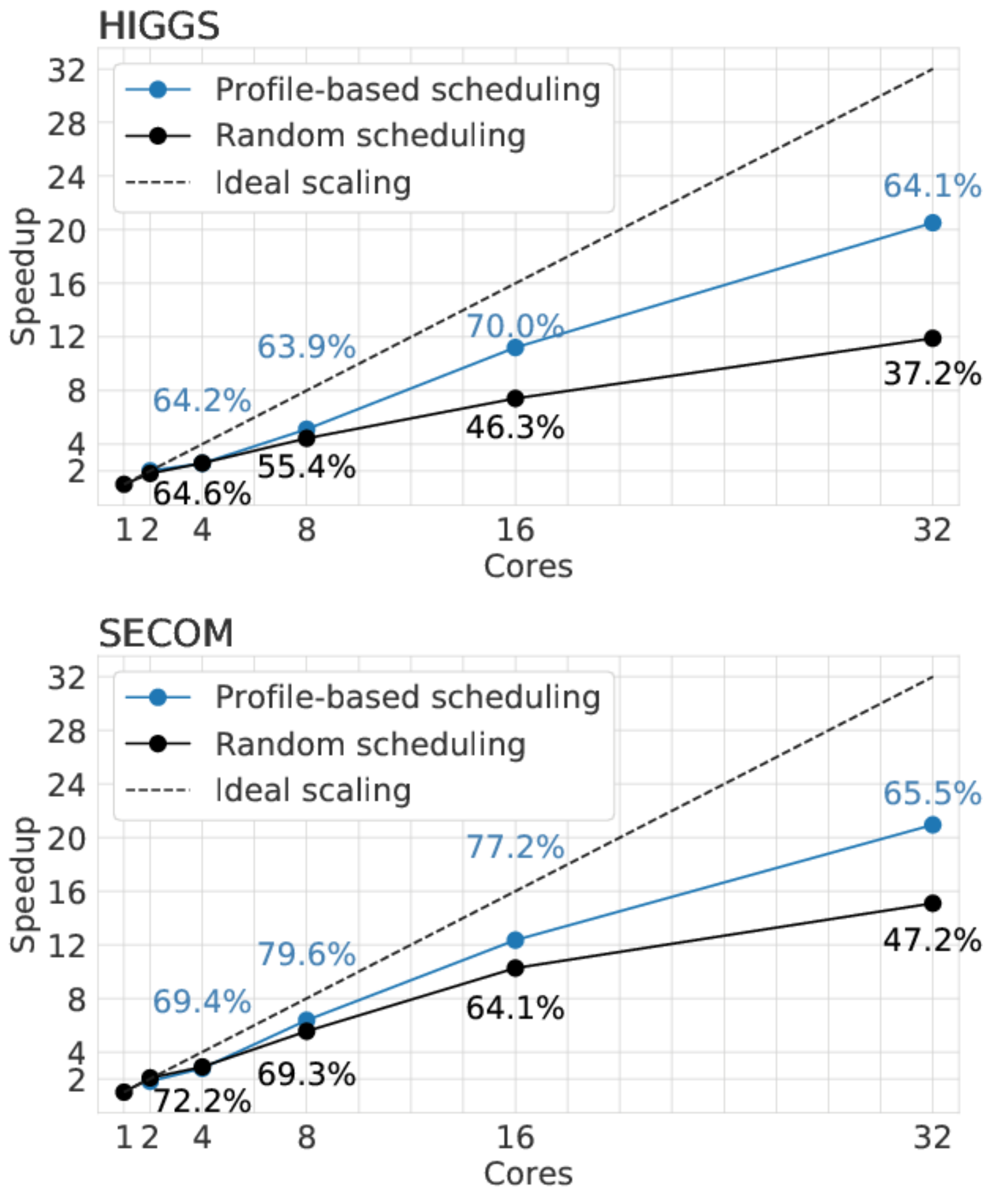}}
 \caption{
Improvements in performance due to increased parallelism.
 }
\label{figure:scaleOut}
\end{figure}

To evaluate our scheduling performance, we compared our proposed profile-based scheduling method with random scheduling for different numbers of cores.
Figure \ref{figure:scaleOut} plots the performance improvements as a percentage of the ideal scaling, based on the results for one core.
Compared with random scheduling, the performance of our profile-based scheduling approach was significantly better when the degree of parallelism was large.
Even though our method requires some profiling time at the start of the search process, it can still improve performance considerably.

\begin{figure}[t]
\includegraphics[width=8cm]{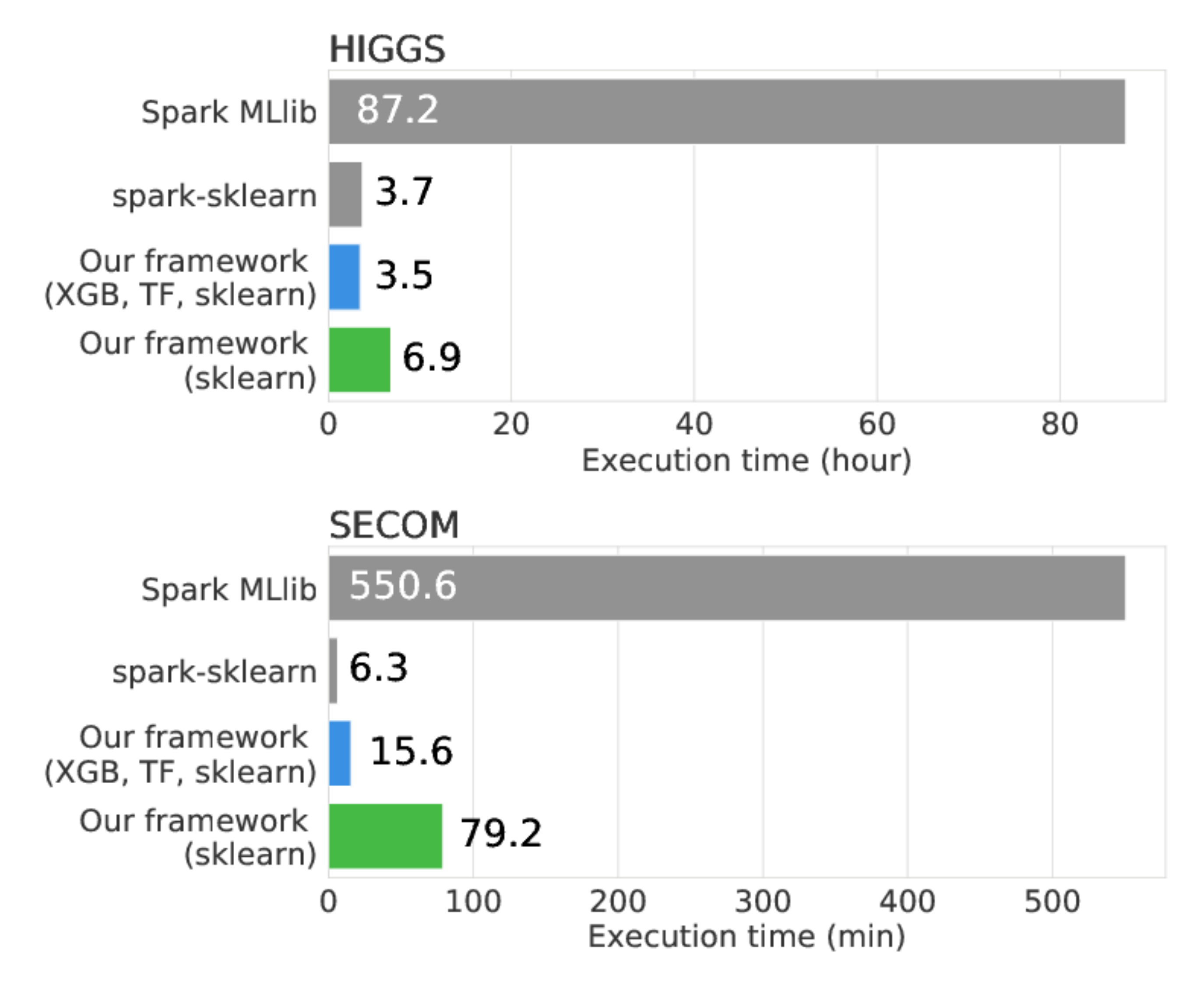}
 \caption{
 Comparison of model search execution times for our framework with those for Spark MLlib and spark-sklearn.
}
 \label{figure:execTime}
\end{figure}

\begin{table}[t]
\caption{
Implementations used to incorporate each algorithm into our framework.
}
\label{table:mapping}
\begin{tabular}{lll}
\hline
 & \begin{tabular}[c]{@{}l@{}}Our framework\\ (XGB, TF, sklearn)\end{tabular} & \begin{tabular}[c]{@{}l@{}}Our framework\\ (sklearn)\end{tabular} \\ \hline
Boosting tree & XGBoost & scikit-learn \\
Multilayer perceptron & TensorFlow & scikit-learn \\
Logistic regression & scikit-learn & scikit-learn \\
Random forest & scikit-learn & scikit-learn \\ \hline
\end{tabular}
\end{table}

Figure \ref{figure:execTime} compares the total model search execution time for our framework with those for Spark MLlib and spark-sklearn.
Here, the colored bars indicate the results for our framework when it incorporated both XGBoost (XGB), TensorFlow (TF), and scikit-learn (sklearn), as shown in Table \ref{table:mapping} (blue bars), and with only scikit-learn (green bars).
These results show that combining all three ML implementations gave better performance than using scikit-learn alone, demonstrating that our framework can benefit from newer implementations with higher performance than traditional ones.

However, comparing our framework (sklearn) with spark-sklearn shows that spark-sklearn achieved better performance, even though they both ran scikit-learn during training.
After investigating this, we found that the performance difference was due to the overhead of generating new Python processes and passing training data to them via the file system.
Thus, there is room for performance improvement if we can either reuse the same Python process or utilize a serialization format with low serialization and deserialization costs.
Even so, despite suffering the same overhead when combining XGB, TF, and sklearn, our framework was still the fastest at conducting a model search for the HIGGS dataset.

Spark MLlib was slower than the other frameworks due to the overhead imposed by the shuffle process needed to train a single model in a distributed fashion.
MLlib distributes the training data among the nodes and manages it, constructing one model by training local models and communicating them between nodes in a shuffle process.

\begin{figure}[t]
\includegraphics[width=8cm]{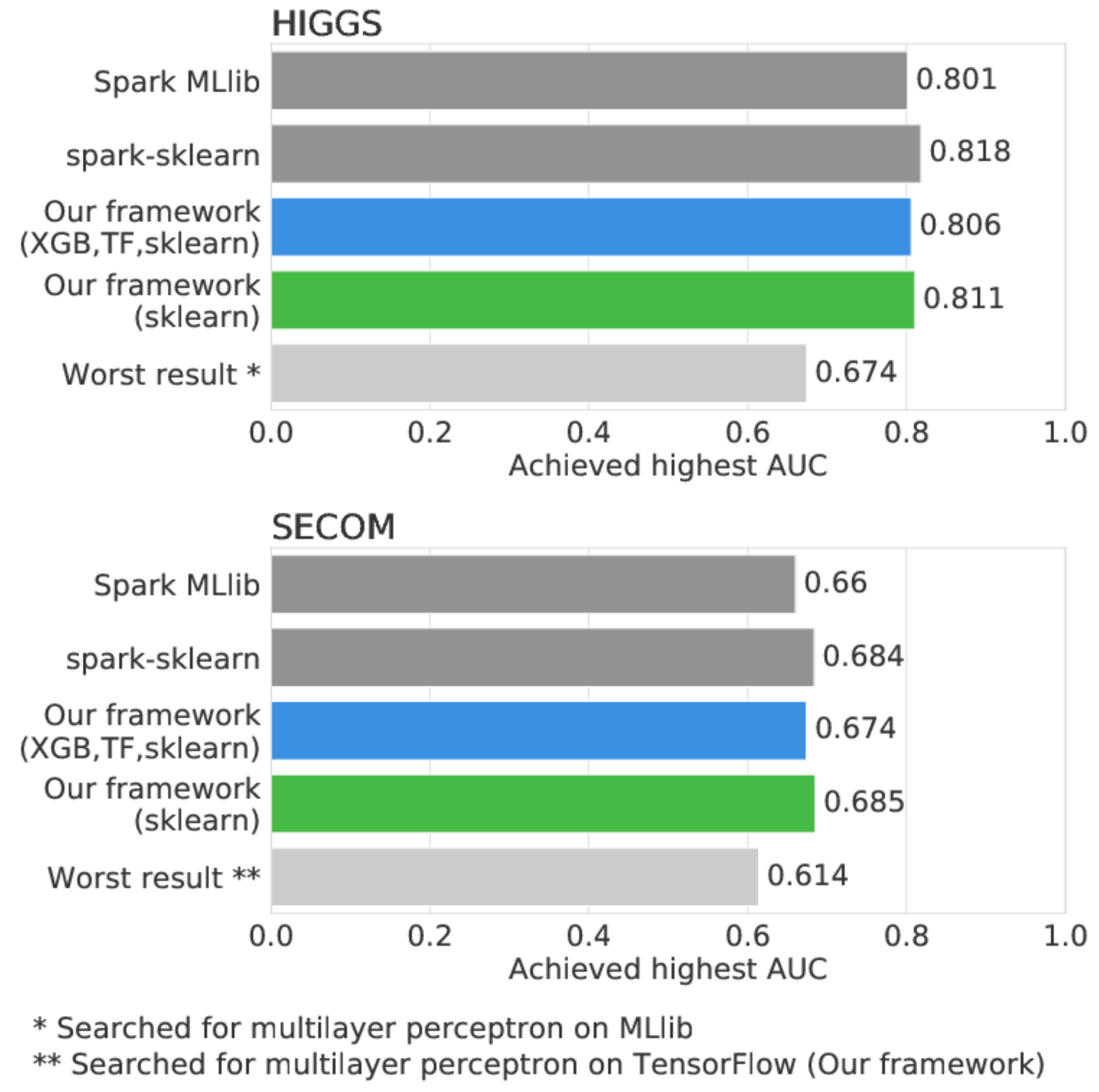}
 \caption{Achieved AUC accuracy for the three frameworks: Spark MLlib, spark-sklearn and our framework. "Worst result" shows the case of searching only one ML algorithm for the model search.}
 \label{figure:accuracy}
\end{figure}

Figure \ref{figure:accuracy} shows the area under the ROC curve (AUC) accuracies achieved by all the frameworks. In addition to the results quoted previously, this also shows the worst results achieved in each experiment, which were for investigating the MLP algorithm, using MLlib for the HIGGS dataset and TensorFlow for the SECOM dataset.
Comparing the other results with the worst ones shows that all the frameworks achieved better accuracies by considering multiple algorithms than with only a single algorithm.

Figure \ref{figure:accuracy} also confirms that all the frameworks achieved almost the same accuracy, and thus executed model search correctly.
Since all four results involved the same set of algorithms and our framework depends on external implementations, it is sufficient to ensure that all the frameworks achieved almost the same accuracy.

\section{Conclusion}\label{conclusion}
In this paper, we have proposed a distributed framework for performing model search across multiple ML implementations.
Our framework has two attractive features: i) new ML implementations can be added by writing common interfaces between the framework and the ML implementations and ii) it scales well with the degree of parallelism due to its profile-based scheduling approach.

We have also shown that, with our framework, implementers need only write 55-144 lines of code to add a new ML implementation.
Additionally, our framework achieved model search results that were the fastest for the HIGGS dataset, and second-fastest for the SECOM dataset.

In future work, we plan to reduce the overhead caused by adding ML implementations in different languages to our framework.
The process invocation overhead could be improved by reusing the same process once called, and the overhead due to passing training data via files could be improved by using a format, such as Apache Arrow, that has low serialization and deserialization costs.

Another problem worth investigating is to assess how the overall execution time is affected by the data sampling rate used for profiling and the number times the profiling step is executed.
Increasing either of these would increase the overall execution time due to the profiling time required, but reducing either of them significantly could adversely affect scheduling performance.
Thus, we also plan to study the profiling parameters needed to minimize the overall execution time in future work.

\section*{acknowledgment}
This work was supported by New Energy and Industrial Technology Development Organization (NEDO).


\ifACM
\bibliographystyle{ACM-Reference-Format}
\else
\bibliographystyle{IEEEtran}
\fi

\bibliography{reference}

\end{document}